\documentstyle[12pt,psfig,epsfig]{article}
\renewcommand{\topmargin}{-0.8cm}
\newcommand{\be}{\begin{equation}}
\newcommand{\ee}{\end{equation}}
\newcommand{\bef}{\begin{figure}}
\newcommand{\eef}{\end{figure}}
\newcommand{\bb}{\bibitem}

\begin{document}
\vspace*{2cm}
\noindent
{\Large {Cosmological QCD Phase Transition and Dark Matter}}\\

\noindent
Abhijit Bhattacharyya$^a$, Jan-e Alam$^{a,b}$, Sourav Sarkar$^a$, 
Pradip Roy$^c$,

\noindent
Bikash Sinha$^{a,c}$, Sibaji Raha$^d$ and Pijushpani Bhattacharjee$^{e}$\\

\noindent
{\small {\it a) Variable Energy Cyclotron Centre,
     1/AF Bidhan Nagar, Calcutta 700 064, India}}

\noindent
{\small {\it b) Physics Department, Kyoto University, Kyoto 606-8502,
Japan}}

\noindent
{\small {\it c) Saha Institute of Nuclear Physics,
           1/AF Bidhan Nagar, Calcutta 700 064,
           India}}

\noindent
{\small {\it d) Dept. of Physics, 
	        Bose Institute, 
		93/1, A.P.C. Road,
                Calcutta 700 009,
           India}}

\noindent
{\small {\it e) Indian Institute of Astrophysics,
		Bangalore - 560 034,
           India}}\\

We calculate the size distribution of quark nuggets, which could be formed
due to first order QCD phase transition in the early universe. 
We find that there are a large number of stable Quark Nuggets which 
could be a viable candidate for cosmological dark matter. 

\section{Introduction}

A first order quark-hadron phase transition in the early universe 
at a critical temperature 
$T_c\sim 100-200$ MeV, lead to the formation of 
quark nuggets (QN), made of $u$, $d$ and $s$ quarks \cite{witten}.
Under certain circumstances the primordial QN's will survive till 
the present epoch. The central theme of this work is the candidature of
these quark nuggets as the baryonic component of the dark matter.
This possibility leaves the results of big bang nucleosynthesis
unaffected and does not invoke any exotic physics~\cite{alam,yang,rana}.

One of the significant issues in this context is the stability of these 
primordial QN's on a cosmological time scale. This question was first  
addressed by Alcock and Farhi~\cite{alcock1} who argued that 
due to baryon evaporation from the surface, a QN, even with the largest 
possible baryon number, will not
survive till the present epoch. Madsen {\it et al}~\cite{madsen1} then 
pointed out that as the evaporation makes the surface of the nugget 
deficient in $u$ and $d$ quarks, further evaporation is suppressed.
They came to a conclusion that QNs with initial baryon number 
$N_B\ge 10^{46}$ could well be stable against baryon evaporation.
Later Bhattacharjee {\it et al}~\cite{pijush} found, using the chromoelectric 
flux tube model, that QN's with baryon number larger than $10^{42}$, 
would survive against baryon evaporation.

In spite of these efforts, not much emphasis has been put 
towards the study of the size-distribution of the QNs.  
The size-distribution of the QNs is very important in the context 
of their candidature as dark matter
as it tells us the most probable size of a QN.
The calculation of the lower cut-off in size
tells us the minimum size and the baryon number content
of a QN that we should  look for. 
We will carry out these studies in the cosmological QCD phase 
transition scenario. 

\section{The size-distribution of quark nuggets}

The evolution of the cosmological scale factor during the quark-hadron phase
transition epoch is given by,
\be
R(t)/R(t_i)=\left(4r\right)^{1/3}\left[cos\left[{{3\left(t-t_i\right)} \over
{2t_c\left(r-1\right)^{1/2}}} +
cos^{-1}{1\over{2r^{1/2}}}\right]\right]^{2/3}
\ee
where $r = g_q/g_h$, $t_c = \sqrt{3m_{pl}^2/8\pi B}$, 
$t_i$ is the time when phase transition starts and $B$ is the bag
constant. (For details and
explanation of all the terms see ref. \cite{abh}).

In the coexisting phase, the temperature of the universe remains constant at
$T_c$. In the usual picture of bubble nucleation in a first order
phase transition scenario hadronic matter starts appearing 
as individual bubbles. With the progress of time, more and more hadronic
bubbles form, coalesce and eventually percolate to form an infinite network 
of hadronic matter which traps the quark matter phase into finite domains. 
The time when the percolation takes place is usually referred to as the 
percolation time $t_p$, determined by a critical volume fraction 
$f_c$, ($f_c \equiv f(t_p)$) of the quark phase.

In an ideal first order phase transition, the fraction of the high
temperature phase decreases from the critical value $f_c$, as these domains
shrink. For the QCD phase transition, however, these domains should become
QN's and as such, we may assume that the lifetime of the
mixed phase $t_f\sim t_p$. The probability of 
finding such a domain of trapped quark matter of
co-ordinate radius $X$ at time $t_p$ with nucleation rate $I(t)$ is 
given by \cite{kodama},
\be
P(X,t_p)=\exp\left[-\frac{4\pi}{3}\int_{t_i}^{t_p}dtI(t)R^3(t)\left(X
+X(t_p,t)\right)^3\right]
\ee
where $X(t_p;t)$ is the coordinate radius of a bubble, at time $t_p$, which 
nucleated at time $t$. 

For convenience, we define a new set of variables 
$z=X R(t_i)/vt_c$, $x=t/t_c$ and $r(x)=R(x)/R(x_i)$; $v$ is the radial 
growth velocity of a bubble. Then 
\be
P(z,x_p)=\exp\left[-\frac{4\pi}{3}v^3t_c^4\int_{x_i}^{x_p}dxI(x)\left(zr(x)
+y(x_p,x)\right)^3\right]
\ee
where
\be
y(x,x\prime)=\int_{x\prime}^x{r(x\prime)}/{r(x\prime\prime)}
dx\prime\prime
\ee

In order to find the minimum size and the size-distribution of the 
QNs we follow the procedure of Kodama {\it et al} \cite{kodama}. 
The distribution function, in terms of $z$, is given by\cite{abh,kodama}
\be
F(z) = {{R^4(t_i)} \over {v^4 t_c^4}} f(z) 
\ee
where
\begin{eqnarray}
f(z) &=& {{3 {\hskip 0.04in} \theta(z-\alpha)} \over 
{4 \pi \alpha^3 }} 
\left[-P'(X-\alpha) -{{3P(X-\alpha)} \over \alpha}\right. \nonumber\\
&+& \left.{1 \over \alpha^2} \int_0^\infty d\eta P(\eta+X-\alpha) 
\left\{\lambda e^{(-\lambda \eta/\alpha)} 
+\omega e^{(-\omega \eta/\alpha)} 
+{\bar{\omega}} e^{(-\omega \eta/\alpha)}\right\}\right]\nonumber\\
\end{eqnarray}

The solution of the equation $F(\alpha) = 0$ gives the minimum size,
$\alpha$, of the quark-nugget. The number of nuggets per unit volume is
then 
\be
n_Q = R^{-3}(t_p) \int_\alpha^\infty F(X) dX 
 = R^{-3}(t_p) \int_\alpha^\infty {{R^3(t_i)} \over {v^3 t_c^3}} 
 f(z) dz
\ee

The volume of each quark nugget is given by 
${4 \over 3} \pi (zvt_c)^3$. Since visible baryonic matter 
constitutes only ten per cent of the closure density
($\Omega_B=0.1$ from standard big bang 
nucleosynthesis), a total of $10^{50}$ baryons will
close the universe baryonically at $T=100$ MeV. We emphasize at this point that
these QNs would not disturb the standard primordial nucleosynthesis
results.
Therefore, if we assume that the total baryon content of the dark
matter is carried by the quark nuggets then,
\be
N_B = 10^{50} (100/T (MeV))^3 = V_H {{4\pi R^3(t_i)} \over {3
R^3(t_p)}}\rho \int_\alpha^\infty f(z) z^3 dz
\ee
where $V_H$ is the horizon volume and $\rho$ is the baryon density
inside each nugget. We have taken $\rho = 0.15 fm^{-3}$ and $v = 0.5$ in
the present calculation. The above equations are then solved self-consistently 
to obtain $\alpha$ and $t_p$. These values are then used to study
the size-distribution of the quark nuggets. To calculate the 
size distribution of QNs we have used
the nucleation rates proposed by Csernai and Kapusta~\cite{kapusta}.

\bef
\centerline{\psfig{figure=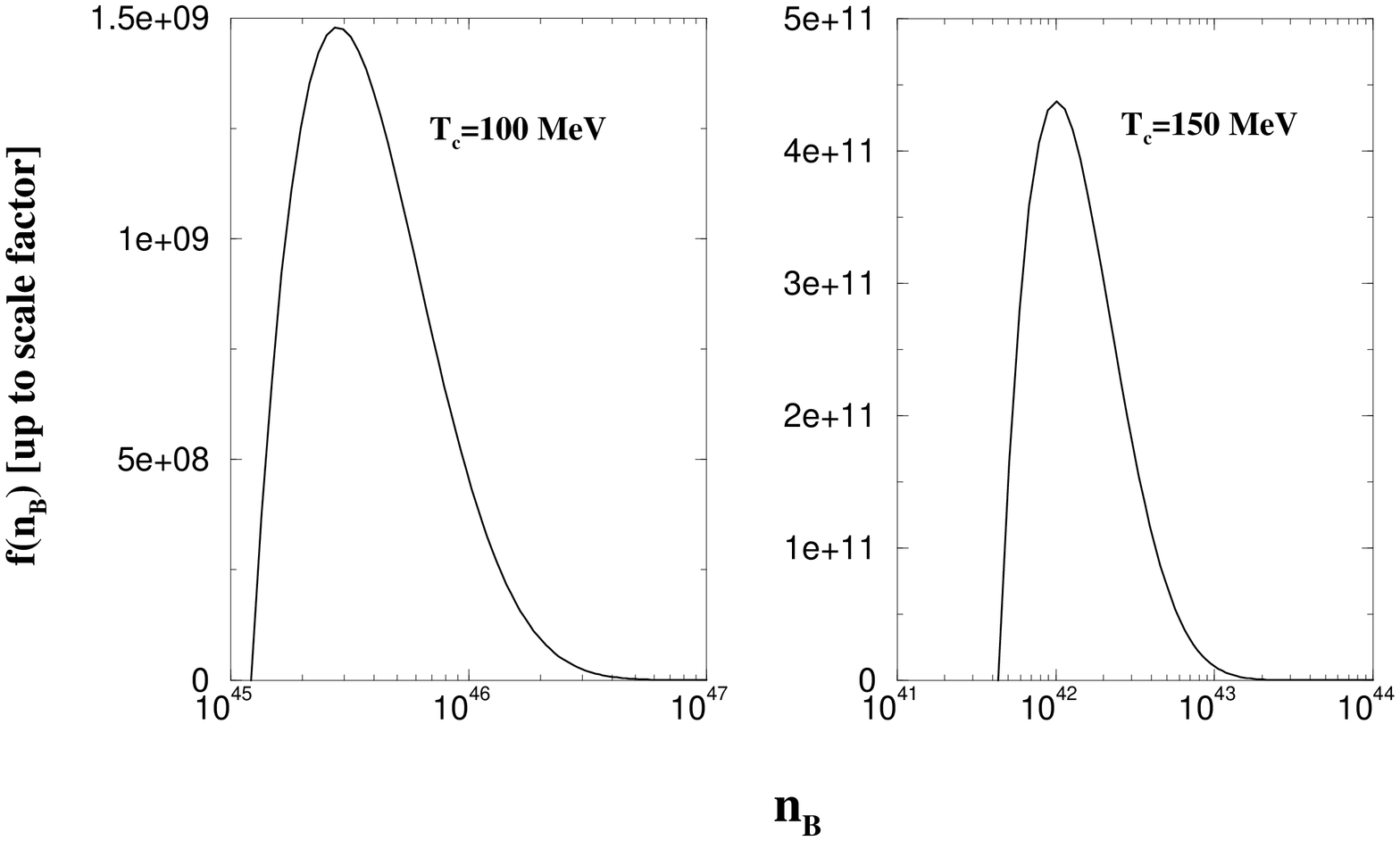,height=6cm,width=14cm,angle=-90}}
\caption{Distribution of QN, $f(n_B)$, as a function of $n_B$ using 
nucleation rate proposed by Csernai and Kapusta. The value of surface
tension is 
$50 MeV fm^{-2}$.}
\label{10kap}
\eef

Once the baryon density inside the
nuggets is known one can easily translate from $z$ to $n_B$ (baryon
number of a particular QN). 
In fig.~\ref{10kap} we have plotted the distribution of QN, 
$f(n_B)$, as a function of $n_B$. 
We see that for $T_c = 100 MeV$ there
is no quark nugget below $n_B = 10^{46}$ and above $n_B = 10^{47.5}$. 
For $T_c = 150 MeV$ there is no quark nugget below $n_B = 10^{41.5}$ and 
above $n_B = 10^{43.5}$. 
Earlier studies \cite{pijush} have shown that the nuggets having baryon
number less than $10^{42}$ will not survive till the present epoch. 
So all the nuggets for $T_c = 100 MeV$ will survive and some of the
nuggets for $T_c = 150 MeV$ will survive.

\section{Conclusion}
In this work we have estimated the abundance of quark nuggets in 
various nucleation scenarios with different values of critical
temperature. We have found 
that within a reasonable set of parameters QNs may be a possible
candidate for cosmological dark matter. 

One of us (JA) is grateful to Japan Society for Promotion of 
Science (JSPS) for financial support.

\end{document}